# Modal Approaches in Metaphysics and Quantum Mechanics

Vladislav Terekhovich[1]

**Abstract.** Some interpretations of quantum mechanics use notions of *possible states* and *possible trajectories*. I investigate how this modal approach correlates with several metaphysical conceptions of a transition from potential to actual existence. The comparison is based on a discussion in contemporary analytical metaphysics of modality. I also consider an analogy between Leibniz's theory of the possibilities striving towards existence, wave mechanics of Schrödinger, and Feynman path integral.

## 1 Introduction

In many interpretations of quantum mechanics (QM), histories and states of quantum particles are directly compared with *possibilities* (Fock, 1957; Heisenberg, 1987; Everett, 1957; Bohm, 1980; Van Fraassen, 1991; Dieks, 2007; Lombardi and Castagnino, 2008; Suárez, 2011). In other interpretations of QM, the modal notions of *possible existence* and *actual existence* are often implicit, although the authors do not give them a special attention. According to Bub (1997), with hindsight a number of traditional interpretations of quantum theory can be characterized as modal interpretations. In particular, these are the Dirac-von Neumann interpretation, Bohr's interpretation, and Bohm's theory.

Indeed, the metaphysical model that divides the reality into possible and actual realms is very popular in QM. One of the reasons is that this model facilitates a conversation about unusual quantum phenomena. For example, the quantum probability is more convenient to be referred not to a set of actual events, as in statistical physics, but to possible outcomes of evolutions of a single quantum object. The alternative states within a quantum superposition, which are incompatible with the classical world, it is more convenient to call the possible states. Accordingly, the quantum world (or worlds) can be presented in the different interpretations as the set of possible states (histories) of quantum particles (fields) or as a possible proposition about them. While the world of classical objects and unambiguous laws, it is more convenient to call the actual world. Dividing of the reality into two realms allows us to alleviate the contradictions between quantum phenomena and habitual ideas about the nature. Greenstein and Zajonc (2008, p. 205) suggests the quantum superposition as the coexistence of possibilities in which interference may occur, whereas a classical picture of mixed states is represented as a set of actual states for which interference is not possible.

[1] Saint Petersburg University, Russia, v.terekhovich@gmail.com

It is known that the possible and actual realms of reality or the possible and actual modes of existence are subjects of several areas of philosophy. At the same time, modern physicists often create their physical models using the concepts of "potential", "possible", and "actual". They do it rather intuitively and without proper criticism sometimes even in the everyday sense. Indeed, we rarely think what the transition from a possible existence into actual means, and why this was borrowed from philosophy. Gradually, we have a situation when any applications of achievements of philosophy are farmed out personal preferences of particular physicists that protect one of the particular interpretations of QM.[2] We may assume that it is quite a natural result since a purpose of an interpretation of any scientific theory is not only to find out a physical meaning of its models, notions, and mathematical symbols. The purpose is also to explain which of these are fundamental, which are not, and how all these are related to the reality and observers.[3]

The main branches of philosophy dealing with problems of the reality, including a possible reality, are ontology and metaphysics.[4] In modern analytic philosophy, metaphysics investigates the issues of being and non-being, existence and reality, universals and individuals, possibility and actuality, God's existence, causality and free will, consciousness and matter, space and time (Loux and Zimmerman, 2005). The first four pairs of these issues are often considered within a framework of ontology. The parts of philosophy exploring the possible and actual realms of reality sometimes are called modal ontology or metaphysics of modality. There they understand under modalities: possibility (it may be) impossibility (it cannot), necessity (it must be), and contingency (it may be or not).

The theme of possible reality has a rich history in philosophy (Gaidenko, 1980; Losev, 1993; Semenyuk, 2014). If ontology is rather a science about essence how essence and about the most general and abstract principles of existence, modern metaphysics, particularly within an analytical approach, is more interested in specific manifestations of the essence into particular things and phenomena. Metaphysics of modality has held a special importance since the 1960s when the semantics of modal logic (Hintikka, 1970; Kripke, 1980). In contemporary metaphysics, among the central themes are problems of possible and actual existence of things or events, pure possibilities (possibilia), and possible worlds (Horuzhy, 1997; Bird, 2006; Yagisawa, 2014). In metaphysics modality, among other things, they study the difference between concepts of "*be*", "*exist*", and "*be actual*". They also are finding out why only some of possible events become actual, and what happens to unrealized possibilities. In theories of possible worlds, they examine how these worlds are connected with each other and the actual world (Adams, 1974; Lewis, 1986; Fine, 1994; Divers, 2002; Armstrong, 1997, Possible Worlds, 2011). A special discussion is devoted to distinguishing the kinds of possibility and necessity (Fine, 2002; Vaidya, 2011). Some authors consider the link between probabilistic causality and theory about the implementation of propensities or dispositions (Kazyutinsky, 2006; Popper, 1990; Suárez, 2011).

---

[2] A description of some attempts see: Petchenkin, 1999, 2000; Sevalnikov, 2006, p. 129; Wilson, 2006.

[3] About the role of models and interpretations in science see: Stepin, 1999; Frigg and Hartmann, 2012.

[4] These terms of fields of philosophy are often perceived as synonymous. In European continental and Russian philosophical tradition, the term of metaphysics is used more in a historical context. The English-language analytic philosophy initially denied metaphysics, but eventually has returned to it and today considers metaphysics as research of structure and content of the reality and as a way how we talk about this reality (Makeeva, 2011; Loux and Zimmerman, 2005).

There are several ways to use the achievements of modern metaphysics within specific sciences (Mumford, 2013). According to the realistic approach, metaphysics do not tell us about things that actually exist, but about things that may exist and are not detected in observations yet (and perhaps cannot be detected), or about things that might be of the observations. From a point of view of a pragmatic approach, the metaphysician studies the current situation in physics and based on analysis of past achievements and philosophy puts forward the hypothesis. Then he compares the hypotheses that are most suitable at the moment without insisting on anything. In this paper, we will use the pragmatic approach and try to compare perspectives of the possibilities in some interpretations of QM with the conceptions of modalities in metaphysics.

The structure of the paper is as follows. In Section 2, we briefly describe the structure of theories of QM: ontology of theory, a model, a mathematical formalism, an experiment, and their interpretation. The differences between physical and philosophical interpretations are highlighted. In Section 3, we list only several historical examples of metaphysical concepts about a transition from potential or possible to actual existence. We also consider the basic directions of discussion in contemporary analytic metaphysics of modality. Section 4 is devoted to the use of the modal approach to QM. We consider the difference between physical and epistemological kinds of possibilities in the explanation of classical and quantum phenomena. Several interpretations of QM are reformulated taking into account the concepts of metaphysics of modality. Section 5 discusses some analogies between Leibniz's theory of the possibilities striving towards existence, wave mechanics of Schrödinger, and Feynman path integral.

## 2 Models and interpretation in quantum mechanics

The theory that describes a behavior of quantum objects contains not only a mathematical formalism, but also some models that are based on a small number of notions and axioms for idealized physical and abstract objects. It is assumed that these objects obey certain logic and interact with each other under common invariable rules. Initially, the model can be built as a hypothesis, and then they write down equations based on the model, and then the mathematical objects are compared with the objects of the model. In the next step, the model and its equations are tested for the ability to predict observations. If the predictions are wrong, then the model or equations (or both) are corrected.

Besides the predictions, there is another equally important function of the theory. It is an explanation. If an experience in most cases corresponds to the predictions, the physicists often want to find out why this model and its formalism are more successful than others. The first natural desire is to use for the explanation the model and its equations themselves. However, such a pragmatic approach does not always help, despite the successful predictions. There are three reasons for this.

First, the models and their formalisms are often borrowed from other physical theories, whose abstract and physical objects "do not work" in a new model. The fact is that a good physical theory except the model and equations includes so-called ontology of theory.[5] The ontology of theory says which objects should be assumed by existing, and suggests a general idea of the surrounding reality. It is known from history, some models of QM have been borrowed from the wave optics, analytical mechanics, and the theory of the electromagnetic field. This happened despite the perspective on the reality in these theories is substantially different. The only argument was a metaphysical belief of the physicists that all laws of nature are united and similar. Second, the formalism of a new theory often uses new abstract objects, including mathematical ones. Despite the analogues with equations of other theories, these objects are difficult explained in frameworks of the older models. In QM, it concerns to the quantum action, the wave function, the phase of probability amplitude, the quantum operators, Hilbert space, the quantum superposition, and entanglement. Thirdly, the same theory can include several models, each with its own formalism[6], and sometimes with its ontology.

Attempts to adapt the old model with its abstract objects, axioms, and rules to explain a new model and a new formalism often lead to paradoxes. In spite of this, it is not necessary for the physicists to change their models, at least, as long as the models do a good "work". It is accepted to formulate the theories of QM in a few very abstract postulates (see Grib, 2013); and sooner or later a need for interpretation of theory arises. The main goals of interpretation can be formulated as follows: (a) to find out which natural phenomena are hidden behind the equations; (b) to describe all parts of the theory in terms of the existence and reality; (c) to explain relationships of these parts with an experiment, including an operation of preparation and measurement. This suggests that any complete theory is a set of theoretical parts (ontology of theory plus models), formalism (mathematical model), experiments, and their interpretation.[7] Indeed, without adequate interpretation we could not be completely sure that our equations were coincided with the observations by chance or were specially slanted in favor the observations. However, only interpretation is not enough, like any part of the theory the interpretation needs to be confirmed by an experiment. If this confirmation is failed, the physicists improve the model or create a new interpretation. The more difficulties are in explanation of connections between all parts of the theory, the more interpretations arise. It seems just that happens in QM and quantum field theory.

---

[5] Initially, this concept was suggested by Quine (2000) for a complete replacement of philosophical ontology in a sense that there is no ontology without theory.

[6] Some authors (Lipkin, 2010; Yourgrau and Mandelstam, 2000) believe that if different formalisms are equivalent mathematically, then these are based on a common model. However, one can show that the formalisms of Newton, Lagrange, and Hamilton in classical mechanics or the formalisms of Heisenberg, Schrödinger, and Feynman in quantum mechanics correspond to the different models see: Feynman, 1968; Polak, 2010.

[7] See more about the use of models in quantum mechanics (Stepin, 2000, p. 248-330).

In the interpretation of the theory, there are two aspects. For models and equations, a physical interpretation is sufficient; the main purpose of this is to find out physical meanings of theoretical notions and mathematical symbols. For this purpose, they are correlated with usual physical objects, such as particles, fields, energy, or space. The ontology of the theory rather needs a philosophical interpretation, whose purpose is to find out clear ontological grounds of language, objects, and equations of the theory. Another purpose is to explain what objects are fundamental and what are not, and how all these are due to the reality and observers. Trying to give such interpretation of the theory, the physicists, often unconsciously, turn to universal philosophical categories: existence, identity, attitude, state, necessity, possibility, actuality; or to universal principles: unity, symmetry, causality, consistency, and others. However, as we know, all of these are the subjects of study of metaphysics or ontology.

Many physicists do not find it helpful to appeal to philosophy. For them, physics is the only reliable source of knowledge about nature, and, therefore, the ontology of physics is equivalent to whole ontology and whole metaphysics. In contrast to naive realism, according to which the majority of observed objects actually exist, the proponents of scientific realism (Chakravartty, 2014) believe that the theory says about things and events that exist and occur in fact. Therefore, we must believe that the theory is true. There are a number of serious objections against this view (Dewitt, 2013): (a) in the past, good many theories turned out to be false; (b) experiments are always limited; (c) in a heart of every theory there are some idealizations and abstractions, including metaphysical, and thus the laws of the theory are valid only for its models.[8]

As a history of QM has shown, a working theory at the same time can have a few models with their formalisms and also a few interpretations. Let us take, for example, three mathematically equivalent formalisms that are based on the different models. Heisenberg's equations are based on the model where physical characteristics of particles are described by the matrix of time-varying coordinates and momenta. Heisenberg (1987, p. 222-223) in general agreed with the Copenhagen interpretation of QM, but he added a metaphysical idea of Aristotle about a transition from potential state to actual one.

---

[8] Van Fraassen (1980) argues that not every theory that well agreed with experimental data (empirically adequate) describes the truth. For example, the concepts of an electron or gravitational field are very convenient, but it does not mean that they exist in reality. Dirac (1980) believed that a good physical model is supported by our ability to use it to calculate that can be tested experimentally. But at the same time he recognized that the real foundation of our faith in a theory does not lie only in experimental evidence, rather beauty of the theory lends credence to it.

The creator of wave mechanics, Schrödinger agreed with an idea of de Broglie that all objects have a wave nature. Using an analogy with the Hamilton-Jacobi equations in analytical mechanics, Schrödinger connected the wave function with the classical action. He did not guess this analogy, since he was firmly convinced that the wave function is associated with the actual wave that carries an electrical charge and unambiguously describes an evolution of a quantum system (Schrödinger, 1976, p. 229-238; Polak, 2010, p. 564).

At the heart of the Feynman path integral is a geometric model of a summation of rotating arrows that symbolise probability amplitude of virtual or possible trajectories of quantum particles. It was found that the probability of quantum events could be found not only by solving the Schrödinger equation, but also by summing (or integrating) all contributions of all possible probability amplitudes, and then squaring this sum (Feynman and Hibbs, 1968, p. 41). Feynman tried to interpret his model using an analogy with the variational principles of mechanics. For this goal, he represented a particle that simultaneously moves along all alternative virtual paths.

Despite the fact that all three formalisms are in good agreement with the experiments, the models, and interpretations, which were used by their creators, are clearly insufficient to explain the nature of quantum probability, the meaning of the wave function, and its "jump" at the moment of measurement. Of course, Schrödinger's wave interpretation is very visual, but it does not explain an observed path of a single particle, just because the waves have to dissipate. It is even more difficult to find some meaning in the summation of the arrows under the method of Feynman.

## 3 Modalities in ontology and analytical metaphysics

The theme of a transition from potentiality to actuality has been investigated by many famous philosophers. Aristotle (Metaphysics, V, 12; XII, 2) believed that everything changes from existence in potentiality or possibility into existence in actuality. In his design, the potentiality or ability (*dynamis*) due to activity (*energeia*) becomes the actuality (*entelechia*). In medieval scholasticism, the notions of dynamis and energeia were translated into Latin as potency and act. According to Aquinas (1969), everything that is in the world passes from potency to act, but what is in potency, that is not yet, and therefore it cannot act. Then Aquinas deduced the necessity of the existence of God. Nicholas of Cusa (1980, p. 167) wrote that reality comes from possibility, and a movement arises from both of them. Thus the movement connects possibility and reality.

Leibniz united two concepts of possibilities. The first is that God has an infinite number of possible worlds. We can be aware of all of them because, according Leibniz, being is inherent in everything that can be thought, but not everything obtains being. Thus, by the will of God, only one most perfect world is actualized. Another Leibniz's concept is devoted to the possibilities of striving towards existence (Leibniz, 1982, p. 234-235). In the last part of the paper, we will look at this concept in more detail. Kant (1994, p. 170) interpreted the possibility and reality as a priori categories of modality: what conforms to the formal conditions of experience that is possible; what conforms to the material conditions of experience that is actual; and the necessary fact is that which relationships to the actuality are defined according to the general conditions of experience. Hegel (1975, p. 318) wrote that the possibility and the contingency are moments of the reality, and the possibility is an internal part of the reality, and the contingency is an external one. According to Hegel, only a realized possibility takes all parameters of existence that the reality possesses. Hartmann (1988, p. 322) called a central mode of being not the reality, but the possibility. Each of the many possibilities in itself is quite defined. Only a fate of realization of the individual possibility is uncertain. The founder of process philosophy, Whitehead (1969) argued that the actual events, which make up the whole world process, are presented as implementations of other things, forming certain potential forms for each actual existence.

The soviet dialectical school continued the tradition of German philosophers and considered the reality of an individual object as its own actual existence with specific qualitative, quantitative, spatial, and temporal characteristics. It was assumed that each individual possibility could be well defined. However, because a material object has many competing possibilities, a change of the object obtains some uncertainty. They suggested using the probability as a measure of the feasibility of a certain possibility (Bransky and Ilyin, 1985).

The questions of the ontological nature of possibilities and possible worlds become particularly popular in science and philosophy after the development of the semantics of possible worlds in modal logics. There they find out a meaning of a proposition as truth or falsity in all possible worlds (Hintikka, 1970; Kripke, 1980). They started to look at the possible worlds like any consistent set of possibilities, possible objects, or possible states of affairs. The analytic philosophers have been paying a special attention to the correlation between being, existence, and reality of possible states of affairs in different possible worlds (Adams, 1974; Loux, 1979; Lewis, 1986; Fine, 1994; Armstrong, 1997; Divers, 2002). To emphasize the difference between modal logic and metaphysics of modality, they often divide modalities into two kinds: *de Dicto* and *de Re*. The first kind consists of the modalities that expressed in a language and logic in the form of characteristics of propositions that differ by a degree of authenticity of the represented states of affairs. The second kind consists of the modalities that inherent in things and phenomena, regardless of our language. In this paper, we are not interested in the language, rather in quantum objects so that we will focus on the second kind of the modalities.

One of the important results of the discussion in metaphysics of modality is a distinction between kinds of possibility and necessity (Fine, 2002; Vaidya, 2011). When people talk about an epistemological possibility, they usually mean that some event or object's state is not contrary to the knowledge of a particular person (the person assumes that it is possible). An epistemic necessity means that the opposite event or state is contrary to the knowledge of the subject (the person is sure). A logical possibility means that event or state is consistent with some system of axioms and rules. A logical necessity directly follows from these axioms in accordance with these rules, which may be not only the rules of classical logic. The metaphysically possible state is possible by virtue of its own essence or true in one of the metaphysically possible worlds. A metaphysical necessity states about essence or truth in all metaphysically possible worlds. For nomologically possible states are sufficient be consistent with general laws of nature; for example, the law that the cause precedes the effect. Nomologically necessary states directly follow from these laws, and the opposite states contradict them. Physically possible states are not contrary to the general physical laws, such as the special relativity or the second law of thermodynamics. Respectively, a physical necessity directly follows from these laws.

It is considered (Vaidya, 2011) that the physical area of possible events and states is narrower than the metaphysical and logical ones. For example, a time travel is logically and perhaps metaphysically possible, but it is physically impossible. Equations of QM allow the simultaneous existence of a particle in alternative states, but this is impossible in classical logic. A correlation between the epistemological possibilities and other kinds of possibilities depends on an answer to a question of whether a scientific theory and its own equations describe only our knowledge or also some side of reality. The scientific realist answers that a theory describes reality. Thus the epistemological possibility is a reflection of the physical and metaphysical possibilities. The proponents of constructive empiricism (van Fraassen, 1980) admit that reality can be very different from the best theory; hence the area of the epistemological possibility may be either narrower or wider than the areas of the physical or metaphysical possibilities. That occurs because, on the one hand, our knowledge is always incomplete; on the other hand, our fantasies are unlimited.

There is a wide-spread opinion that any object is an actual object. From this follows a concept that non-actual possible objects are nothing. There is another conservative view that any object is an existing object. In metaphysics of modalities, all complex views of this can be divided into *realism* and *anti-realism* or *nominalism* (Rescher, 1975; Lowe, 1998; Divers, 2002). Realism considers every possibility as something that exists in reality, regardless of whether we think about the meaning of these notions. So realism represents the possibility an ontological and cosmological category. Anti-realism, on the contrary, denies the existence and reality of possibilities and possible worlds. Both are announced the creation of our mind and existing only as names, fiction or theoretical constructions. In particular, Kripke (1980) argued that the term "possible world" is merely a tool of a language useful to visualise a concept of possibility.

In terms of analytic philosophy, *being* and *existence* are usually used as synonyms. However, when we need to emphasise the difference between them, we should treat *existence* as one of the ways or *modes of being* (Makeeva, 2011). One of the criteria to classify modern conceptions of modality and possible worlds is the correlation between the notions of being, existence, and actuality. Let us consider some of these conceptions.

The *modal realist* or *Lewis's possibilist* (Lewis, 1986) states that an infinite number of possible worlds exist in actuality, and they are just as actual as our world. Our world is just one among many like it. Possible events or objects exist that are not less than actual events or objects. To exist in the world is simply to be a part of it. They introduce the principle of relativity: objects of possible worlds are possible for us, although these are quite actual objects for the inhabitants of other worlds.

For the *classical possibilist*, possible objects and possible worlds are in the ontological sense, so some of them could have existed in the physical world. The only physical world exists as actual, and it consists of actual objects that exist too.

According to the *Dispositional essentialists* (Ellis, 2001; Bird, 2006), the world is, ultimately, merely like a conglomerate of objects and irreducible dispositions. Dispositional properties are, unlike categorical properties, supposed to be properties that are not wholly manifest in the present; thus, they are the ultimate ontological units that explain events. Any object that possesses the dispositional essence of some potency is disposed to manifest the corresponding disposition under stimulus conditions, in any possible world.

The *actualist* (Adams, 1974) denies the reality of possible objects, and he states that everything is exists like an actual thing. There is nothing that is not an actual thing, so the physical existence equals being. Possible worlds are nothing more than fictions, "ersatz" linguistic constructions are created within the actual world; they are abstract states or conditions in which a concrete world could be. Some of the actualists (Plantinga, 1974) attract unactualised individual essences. In other words, every object has an individual essence independent of the object that has it, whether the object is actual or non-actual. One of the versions of actualism—combinatorism—considers possible worlds as a certain sort of recombination of properties and relations of the objects or states of affairs of the actual world (Armstrong, 1997).

Each of these theories faces many difficulties. Realists, for example, cannot explain how the possible worlds are connected with each other, or how the possible world emerge and obtain reality. Anti-realists try to answer the question of how the possible worlds turn into the actual world, or what happens to possible objects or possible worlds that will never become actual. However, despite all these difficulties, there is no doubt that the results of studies in metaphysics of modality can be successfully used in the interpretations of QM. The examples of such usage will be given below.

## 4 Modalities in interpretations of quantum mechanics

One of the lessons obtained from metaphysics of modality is that when we talk about possibilities in quantum phenomena, we must always specify what kind of possibility we mean. Let us consider three examples of a correlation between the epistemological and physical possibilities. First, we say that it is possible that, in a next room, there are ten chairs, and this possibility has a probability of 70%. Here we estimate a degree of our knowledge or, on the contrary, a degree of our ignorance. The assessment is based on our life experience and also on the knowledge of the physical possibility of this fact. From a physical point of view, at the moment, these chairs are either there or not there. It is a necessary physical fact that is independent of our knowledge. When we enter the next room and see the chairs, we will change the degree of the epistemic possibility, but the physical possibility and necessity will not be changed.

Second, when weather forecasters use an ensemble forecast they say that the probability of rain tomorrow is 70%. This means that only 70% of all scenarios designed by computer end with rain. Unlike the first case with the chairs, they estimate a future event, and this assessment is based on the statistics of rain in previous days. From the epistemological and physical perspectives, rain is possible, but the probabilities of these possibilities are different. The reason is we could not account for all factors when we calculate the weather scenarios. In this case, the epistemological possibility cannot be correlated with only a degree of our ignorance since rain tomorrow is not physically necessary.

Third, if we say that the probability to detect a photon at a particular point of a screen is 70%, this means that the calculation of equations for a single photon gave us this amount. If the equations are adequate to experiments, the epistemological possibility to a certain extent can be considered equal to the physical possibility. However, there is a fundamental difference. To test the epistemological possibility, it is necessary to measure a probability distribution by detecting a large number of photons in a series of experiments. The physical possibility and its measure—a probability amplitude—exist even for a single photon, regardless of our knowledge about this possibility and the behavior of other photons. Like in our second case of rain, the epistemological possibility of detecting a photon cannot be correlated with only a degree of our ignorance. But unlike the physical possibility of rain, which is determined by a random chain of future events, the physical possibility of a photon is determined by its current state or, more precisely, by superposition of many possible states.

The different interpretations of QM use different kinds of possibilities. For example, in Bohr's variant of Copenhagen interpretation, epistemological possibility and physical necessity are considered. Fock (1957) preferred physical possibility and necessity. Many-worlds interpretation tries to combine the physical possibility and metaphysical necessity (Saunders, 2010). Most of modal interpretations tend to the metaphysical possibility and the metaphysical necessity.

To make our reasoning clearer, let us introduce a notion of quantum possibilities that is analogous to the physical possibility that is discussed above. Let a quantum event, a quantum state, or a quantum history be possible if they do not contradict the laws of QM. Accordingly, let us use a name "possible quantum world" for a set of possible states and possible histories of quantum particles (fields) that have some common properties.

To demonstrate how the achievements of metaphysics of modality can be used in the interpretations of QM, let us reformulate the basic conceptions of possibility and possible worlds in respect to quantum objects and their states. Then let us reformulate some interpretations of QM in modal terms.

From the anti-realistic point of view, the possible states and the possible histories of quantum particles (fields) exist only in models and formulas, which are derived from our experience as tools of theoretical study of the actual world. Therefore, these do not have an independent life. In the analysis of the possible worlds, such approach is combined with the conception of *actualism*. According to actualism, there is only one actual classical world; the quantum possible world does not exist in any way, it is mere a set of the possible states (histories) of quantum particles (fields) like possible scenarios or states of affairs that could potentially be found in the actual states of the classical objects.

It is interesting that there are not many completely anti-realistic interpretations concerning to the possibility. First of all, they include Bohr's version of Copenhagen interpretation. According to this, it is meaningless to talk about the reality of the possible states of quantum particles before their measurement because these exist only in formulas. The only actual world is created by the measurement, and the "collapse of a wave function" does not describe a change the reality, but a change of our knowledge of the reality. Thus, the quantum and macro objects are described with fundamentally different languages (the principle of complementarity). The square of the summed wave function is interpreted the probability of detecting of an object at a particular point in space-time. In contrast to the classical probability, the quantum probability is not a consequence of an incompleteness of our knowledge. The quantum probability describes objective characteristics of the quantum objects, even though, in experiments, these characteristics are observed in the form of a frequency or a probability distribution of events in multiple experiments.

In many versions of the statistical or ensemble interpretation of QM (Petchenkin, 2004), it is assumed that the probabilistic laws of QM are only convenient ways to describe the unique and unambiguous actual reality. At the time of measurement, there are recorded not individual characteristics of individual particles, but statistical characteristics of the aggregate of the particles. Despite the fundamentally different views of the nature of quantum probability, both the Copenhagen and the statistical interpretations are close to the metaphysical concept of actualism.

If we accept an opposite—realistic position within metaphysics of modality, we should recognize that the possible states (possible histories) of quantum particles (fields) exist independently of our mind. For example, in terms of essential dispositionalism, the possible states of quantum particles are the manifestation of their objective propensities or potencies. In the analysis of the possible worlds, this approach is combined with two concepts. From the point of view of modal realism, the quantum possible world appears a possible set of possible states (histories) of quantum particles that are actual as well as our world for us. The supporters of the conception of possibilism consider the quantum possible world ontologically like a set of potential states (histories) of quantum particles that have not been realized and yet do not exists in our actual world.

There are many interpretations of QM, which implies some degree of existence of the quantum possibilities. We will list only some of them. The interpretations of Heisenberg and Fock are close to the viewpoint of Bohr but differ by the relation to the possible quantum states. Heisenberg believed that mathematical laws of quantum theory can be considered a quantitative formulation of Aristotelian notions of "dynamis" or "potency", and that a notion of "possibility" occupies an intermediate position between objective material reality and subjective reality. He admitted that the quantum-theoretical possibility has partial objectivity. However, if this possibility is interpreted as a measure of frequency, it would have a meaning only regarding to a set of mentally represented events (Heisenberg, 1987, p. 223). In contrast to Heisenberg, Fock believed that the state, which described by the object's wave function, is objective in the sense that it is independent of the observer characteristic of the potential possibilities of one or another result of the interaction between an atomic object and a device (Fock, 1957). According to Fock (1967, p. 179), in an experiment one of the possible outcomes carried out. The outcome is provided by the initial wave function and corresponds to a new wave function. The probability of a particular object's behavior under given conditions is a numerical assessment of the potential possibilities of this behavior.[9] Such a view is consistent with the metaphysical concept of possibilism.

Sevalnikov (2009) develops a similar idea of the poly-ontic or poly-modes model of QM. He examines a concept of "coexisting possibilities" that means the possibility may intersect another one or include it (ibid, p. 98). According Sevalnikov, dynamic and continuous change of the wave function in the Schrödinger equation describes that it happens on the "potential" level, or that does not actually exist and is not yet actualized. Only during a measurement, when "other" (for example, a device) intervenes, there is the implementation or the actualization of the possibility. Thus, the Schrödinger equation describes a border between the possible and the actual levels of being. Classical physics describes the classical world that is the actual world while the mathematical formalism of QM describes the formation of the actual world (Ibid, p. 143).

When Schrödinger developed his wave mechanics, he did not agree with Heisenberg that an observed trajectory of a single particle is the result that one of the possible trajectories transforms to the actual trajectory. Schrödinger explained the actual trajectory by a set or a field of all possible trajectories. According to Schrödinger (1976, p. 229-238), in an infinite number of possible trajectories none of these has an advantage to be implemented in a particular case, all these are equally real. This view is also close to the conception of possibilism, although instead of the implementation of only one possible entangled state, all ones are summed up. It occurs due to resonance or interference of the waves (Ibid, p. 261-284).

Popper's propensity interpretation of QM based on an objective notion of probability like possibility or potency in terms of Aristotle (Petchenkin, 2002). According to Popper (1998, p. 17), the quantum reality is not a reality of actual always available object's properties, but the reality of object's propensities or dispositions to certain behavior. These dispositions are as well real as physical forces or fields. This view is close to the modern metaphysical theory of essential dispositionalism. In Bohm (1952) theory, it is assumed that a hidden and nonlocal in space-time field of quantum potentials objectively exists. This field depends on positions of all particles that at once influence the actual trajectory of the particle. According to the theory of "Holomovement" (Bohm, 1980), any measurement or interaction extracts objects from an entangled state of "undivided wholeness" or "explicate order". Wholeness is not static, but a dynamic wholeness-in-motion in which everything moves together in a flow. We perceive an object as real, although it existed prior to the measurement since the flow is prior to "things" that emerge from the whole of flowing movement. As soon as the measurement or interaction stops, the object dissolves back into the state of the implicate order. If they regard the non-local field of potentials and implicate order as an analogue of the possible mode of existence, then Bohm's idea is largely consistent with the conception of possibilism.

---

[9] More about the difference between the views of Heisenberg and Fock see in Sevalnikov, 2009, p. 120-127.

It is considered that Feynman did not formulate his own interpretation of QM. He had even pointed out that no one was able to find out a mechanism that hides behind the laws of QM (Feynman, 2004, Vol. 3, p. 207). However, he gave many explanations for his path integral formalism. According to him, photons do follow along all possible mutually-exclusive trajectories, and the summation of their probability amplitudes is not empty play in mathematics (Feynman, 2014, p. 49-54). In addition to the experiments with light, this view of the reality of possible trajectories contributed to an analogy with the principle of least action of classical mechanics. In this principle, a particle "feels" all the neighboring trajectories and selects the one along which the action $S$ is minimal. If we forget about all these probability amplitudes, particle indeed moves along a special trajectory—the trajectory for which $S$ in the first approximation does not change. This is the connection between the principle of least action and quantum mechanics (Feynman, 2004, Vol. 6, p. 111-112). If Feynman's alternative trajectories to be considered as possible trajectories along which a quantum object "moves" (of course, there is no any classical trajectory), then the actual trajectory is the mere sum of all possible trajectories, or rather the sum of probability amplitudes of possible trajectories and their complex phases .

In the Consistent or Decoherent Histories interpretation of QM (Gell-Mann and Hartle, 2012), they select from all alternative quantum histories (analogues of Feynman paths) a set of coarse-grained coherent histories that due to decoherence or "entanglement with the environment" do not almost interfere with each other. Thus, the individual histories in the set obey the usual sum rules of the classic probability theory. When the probability distribution for the consistent histories achieves a strong peak, they behave themselves as the quasi-classical histories. Therefore the classical equations of motion can be considered as a limiting case of the quantum laws. The reality of quantum histories before decoherence does not depend on the measurement or an observer; it means that these can be considered as possible histories that turn to the actual ones under certain conditions. The proponents of this interpretation do not use any modal terms. However, this view is close to the conception of possibilism if instead of the one of many histories they examine the whole mechanism of decoherence of the beam of the possible histories.

Zurek (2003) offers the Existential interpretation of QM based on a mechanism of decoherence with the environment. He tries to combine two opposite point of view: (a) the reality is only our knowledge, as in the Copenhagen interpretation; (b) the reality is an ontological entity. Zurek considers alternative states of quantum objects that are in quantum superposition. The ontological features of the actual states are selected only when the superposition principle is "turned off" by environment induced decoherence. The objective existence of the selected states is acquired through the epistemological information exchange with the environment. This exchange of information exists objectively; it is the cause of any changes and interactions. It is supposed that information is not only human knowledge, but the primary entity. One of the possible alternative states leaves quantum superposition not due to the presence of an observer, but due to the mere existence of the possibility to transfer information to the observer.

In the Many-Worlds interpretation of QM (Saunders, 2010), it is assumed that any measurement of quantum particles divides them into many copies, each of which actually exists in a parallel world or a projection of a multiverse. Any copy evolves according to the Schrödinger's equation; and a wave function is an ontological entity. At first glance, this interpretation resembles the conception of modal realism in metaphysics of possible worlds (Lewis, 1986) where all possible worlds exist and are relatively actual. However, the similarity is deceptive like Lewis (Lewis, 2004) had written. In modal realism, possible worlds (universes) develop independently of each other even under different laws. In the Many-Worlds interpretation, each possible world is mere one of all possible alternative histories or branches of the evolution of the single multiverse. These branches have a common source and common laws but are divided because of decoherence with the environment. Multiverse is composed of a quantum superposition of all its own possible branches or quantum worlds. Interestingly, the creator of this interpretation Everett (Everett, 1957) denied any direct analogy with the transition possibility to the actuality that adopted in possibilism.

In recent decades, a family of interpretations has emerged. The authors have directly called them "modal interpretations." They understood any quantum states as descriptions of a collection of the possibilities that possess own probability and exist in the space of possible events. Despite many differences, all of the modal interpretation based on the standard formalism of QM, except for the projection postulate, providing "collapse of the wave function" (Petchenkin, 2000; Lombardi and Dieks, 2014). Van Fraassen (1991) first proposed that a quantum system has two kinds of states: *dynamical state* and *value state*. The dynamical state determines the system's possible physical properties and their probabilities. For an isolated system, the dynamical state evolves according to the Schrödinger equation (in the non-relativistic case) and never collapses in the process of evolution. That is why, they sometimes call the Modal Interpretation one-world version of the Many-Worlds interpretation. The value state represents actual physical properties or values of physical quantities that accurately identify the system at any time. The measurement as well as any physical interaction randomly detects (but does not create, as Heisenberg supposed) one of the possible value state making it actual. Modality is not a result of the incompleteness of description, and the observer does not play any role. The quantum theory is fundamental not only for particles, but for all macroscopic objects. Thus, the quantum formalism does not say what actually happens in the physical world, but it only gives us a list of the possibilities and their probabilities (Dieks, 2007).

From the point of view of Bene and Dieks (Bene and Dieks 2002, Dieks 2010), the properties of a physical system are defined in relation to other physical systems. Such relational descriptions from different perspectives are equally objective and relative to the physical reality that is also relative. In the modal interpretations, modalities are mere convenient tools for the description of the actual world and do not have their existence. Such a view is closer to the conception of actualism since the possible value state exists within the actual world, and we accidentally discover them.

In the Modal-Hamiltonian interpretation of QM (Lombardi and Castagnino, 2008), the Hamiltonian of a system plays a decisive role what the value state will be actual. They introduce ontology with irreducible to each other realm of possibility and realm of actuality. Quantum systems are within the realm of possibility that is not less real than the realm of actuality. They consider propensities as real properties that follow a deterministic evolution independently of which possible facts become actual. The propensities produce definite effects on the actual reality even if they never become actual. Such a view is close to the conception of possibilism because the value states or propensities are realized in the actual world.

## 5 Leibniz's theory of striving possibles and Feynman path integral

There is another example how to use metaphysics of modality for the interpretation of QM. Let us examine some interesting analogies between Leibniz's theory of the possibilities striving towards existence, wave mechanics of Schrödinger, and Feynman path integral. Let alternative virtual paths in the Feynman formalism are the possible trajectories. Then, from the various metaphysical perspectives, the paths may have different relationships to reality. For example, in terms of modal realism, these may exist in other possible worlds. For the possibilist and dispositional essentialist, Feynman paths have some degree of being in our world but do not yet actually exist there. The actualist would say that Feynman paths are fictions having neither being nor existence.

In spite of the fact that most of the theories of possible worlds are based on ideas of Leibniz, his metaphysical system is significantly different from all modern modal theories and deserves a separate study. In his numerous works, Leibniz developed the Aristotelian view of the possibility as potentiality, at the same time he identified the possibility with an ideality that brings him to Plato (Mayorov, 1973). For Leibniz (1984, p. 124), actuality is something that expresses the existence, but potentiality expresses only the essence. Since Leibniz was a scientist no less than a philosopher, he was not satisfied with too abstract Aristotelian model of the implementation of the potentiality (dynamis) through activity (energeia) to the actuality (entelechia). Leibniz (1982, p. 234-235) tried to imagine how this metaphysical process manifested itself in physical processes. There was a similar distinction between essence and existence. Leibniz postulated that the principle of governing essences is that of possibility or non-contradiction. He suggested (Ibid, p. 283-284) that every essence (possible thing) tends of itself towards existence, but the one that actually exist is that which has the greatest perfection, or degree of essence, or the greatest number of possibles at the same time. The more perfection, the more existence. According to Leibniz, the things are incompatible with the other things; therefore some possible things do not achieve their actualization. Thus, from the collision of all possibilities, only those things will be actualised that contain the greatest number of the possibilities. In other words, "the possibles vie with one another for existence by combining forces with as many other essences as they are mutually compatible with" (Blumenfeld, 1973). So the world arises in which the largest part of the possible things is actualized (Leibniz, 1982, p. 285). Leibniz gave an example of such things: a straight line among all lines, a right angle among all angles, a circle or a sphere among all figures as the most capacious ones.

Let us return to quantum phenomena. As we already mentioned, Schrödinger (1976, p. 229-238) tried to explain the observed trajectories of a particle by a complete set of possible trajectories, rather than the transformation of one of them into the actual trajectory. On the other hand, we could rephrase foregoing Feynman's statements concerning his path integral formalism. These are as follows: the summing of probability amplitudes of all possible trajectories entails the realization of the trajectory that integrates the largest number of probability amplitudes with close phases. The observed trajectory differs from others with the maximal probability.

In this table of correspondence, we bring together some notions from the conceptions of Leibniz, Schrödinger, and Feynman.

| Metaphysics of Leibniz | Quantum Mechanics |
| --- | --- |
| Amount of existence | Square of amplitude of probability |
| Measure of necessity of individual possibility | Probability |
| Collision or competition of possibilities | Interference or summation of probability amplitudes |
| Coexisting or compatible essences | Superposition of coherent trajectories |
| Maximal degree of existence | Observed trajectory |

Thanks to this table of correspondence, we can translate some metaphysical conclusions into the language of QM. For example, the statement that not actualized possibilities always compete and join with other compatible essences, seems very similar to the statement: "The alternatives that cannot be detected in the experiment always interfere with each other" (Feynman and Hibs, p. 26). There are other examples of amazing correspondence between the statements based on metaphysics of Leibniz and the language of QM.

*Metaphysics*: possible histories have the essence and are in a possible mode of being, but these do not yet exist in the actual world since these are incompatible there. *QM*: before the interaction or measurement, the evolutions of alternative trajectories of particles are described by unambiguous wave functions, but these trajectories do not exist as real ones.

*Metaphysics*: each possible history has its own measure of necessity with which this history becomes actual. *QM*: the probability amplitude (and its complex phase) of the specific alternative trajectory of the particle is related to the probability of observation it at this trajectory.

*Metaphysics*: within the possible mode of being, an object moves at once along the infinite set of all possible histories that are compatible in the space of possible events. The possible histories of the specific object are compatible with each other, and therefore they are joined. *QM*: in Hilbert space, the particle simultaneously moves along all alternative trajectories that are in a coherent superposition. Their probability amplitudes are summed up; in the wave representation, it looks like interference.

*Metaphysics*: after the collision, competition, and combination between the possible histories, the result has the maximal essence and manifested in the actual existence as the unique history. Other possibilities remain in the possible mode of being. *QM*: the resulting trajectory of the particle with the maximal probability is the unique one in smooth four-dimensional space-time. Other alternative trajectories continue to take part in the quantum superposition. Since their contributions to the resulting trajectory are relatively small, they are not observed.

These analogies can be applied not only to the quantum trajectories but also to the quantum states. Each quantum object can be represented as the superposition of all its own possible alternative states. The transition from the set of possible states to a single actual state should occur after each interaction between the object and its environment, including both quantum and macroscopic objects. Summing up, we can assume that the totality of all possible motions and possible states of quantum objects forms the possible realm of reality, and the set of the actual motions and actual states forms the actual realm of reality. These realms are "parallel" and continuously pass into each other. This conclusion is consistent with the common metaphysical hypothesis that quantum systems belong to the sphere of potentiality, and classical systems belong to the sphere of actuality.

## 6 Conclusion

The interpretations of QM in which it is assumed that quantum states describe the possibilities can be divided into four groups. In the first group, one possibility becomes actual due to observation; in other words, the fact of the observation creates the reality from the possibility. This is the approach of the Copenhagen interpretation that was particularly clearly formulated by Heisenberg and Fock, who, in fact, followed Aristotle.

In the second group of interpretations, they assume that each possible state is realized as actual one. The fact of observation or the choice of consciousness selects one of the states in one of the worlds (in one of the branches of the multiverse). This point of view is supported in the Many-Worlds interpretation.

The third group unites the various Modal interpretations of QM. These argue that one of many possible states is only detected but not changed by observation. In one of the versions of this interpretation (Dieks, 2010), the possibility is reduced to the reality that corresponds to the conception of actualism in metaphysics. In another version (Lombardi and Castagnino, 2008), they accept the possibilistic conception that possible events constitute a fundamental ontological category. The probability is considered as an objective measure of propensity of some possible quantum event to become actual. The second version does not deny the former.

According to the fourth group, the entire set of the possible states of a system "observer-device-object" is realized. However, we observe only a cumulative result of their summation. The actual reality is the mere sum of all possibilities that coexist in the potential reality of quantum objects. The similar ideas were expressed by de Broglie and Schrödinger. The same conclusion can be reached through the analogy between Leibniz's metaphysical theory of possibilities striving towards existence and the Feynman path integral formalism. The hypothesis of "summation of coexisting possibilities" (Terekhovich, 2012, 2013) applies a similar approach not only to quantum objects but also to classical ones.

The diversity of views of the possible quantum states shows how productive can be a comparison between the modal approaches in the interpretation of QM and an analysis of possibilities in metaphysics. We assume that the achievements of metaphysics will be useful in the investigation of the specific interpretations of QM as well as in the comparison of their heuristic value. In addition, this will take a fresh look at many old problems of both physics and metaphysics.